\documentclass[11pt]{article}
\usepackage[super,compress]{cite}
\usepackage{graphicx}
\usepackage{amsmath}
\usepackage[colorlinks=true,linkcolor=blue,citecolor=blue,urlcolor=blue]{hyperref}

\usepackage{ulem}
\usepackage{ragged2e}  % turn back to justifying paper style

\usepackage{soul}

\date{}
\author{
B. Dudás$^{1}$, G. Papp$^{1}$, G. G. Barnaföldi$^{2}$, G. Bíró$^{2}$\\
for the Bergen pCT Collaboration\\[1ex]
$^1$Institute for Physics, Eötvös Loránd University,
\\ Pázmány Péter sétány. 1/A, 1117, Budapest, Hungary\\
$^2$HUN-REN Wigner Research Centre for Physics,\\
29--33 Konkoly--Thege Miklós út, H-1121 Budapest, Hungary\\[1ex]
{\small \texttt{dudas.bence@ttk.elte.hu}}
}

\title{Compact deep learning pipeline for particle track reconstruction in the pCT detector system}
\begin{document}

\maketitle
\justifying

\begin{abstract}

Proton computed tomography (pCT) requires both fast and accurate reconstruction of particle trajectories and kinetic energies to achieve clinically viable image formation. Traditional distance-based matching algorithms often fail under the combined effects of multiple Coulomb scattering and track crossings and most importantly many of them take too much computation time, motivating the use of lightweight deep learning models that can be evaluated rapidly. In this work, we develop a two-stage reconstruction pipeline consisting of (i) a neural-network-assisted tracking module and (ii) a kinetic-energy estimation model. For the tracking task, compact 
multilayer perceptrons are trained to predict the expected hit position in the subsequent detector layer, providing a physically informed prior that substantially reduces ambiguities in bipartite matching. Furthermore, ambiguous tracks are flagged and excluded from the final analysis.

Our training data is  provided by OpenGATE simulation toolkit, both for tracking and energy estimation, 
where we designed a fully connected network that processes detector hit information. 

This model predicts the incoming proton kinetic energy with sufficient accuracy for current pCT image reconstruction methods. The entire pipeline benefits from deep-learning parallelism and evaluates particle tracks fast enough for clinical time constraints.

Together, these results demonstrate that compact deep learning models can reliably reconstruct particle trajectories and energies in a realistic pCT detector system, offering a computationally efficient and highly accurate alternative to traditional matching and tracking methods.
\end{abstract}

\section{Introduction}
The research presented in this paper aims to contribute to Proton Computed Tomography~\cite{pctarticle} and by that the advancement of hadron therapy~\cite{wilson1946radiological}, a form of radiation-based cancer treatment that employs protons and heavier ions such as helium, carbon, and oxygen to achieve greater dose concentration compared to conventional X-ray therapy. A key advantage of hadron therapy arises from the characteristic energy loss behavior of charged particles in matter. 
In contrast to photons (e.g., X-rays), which deposit a substantial fraction of their energy upon entering the tissue, the energy deposition characteristics of charged particles are defined by the Bragg peak~\cite{Braggpaper}. This phenomenon ensures that the majority of the energy is released at a well-defined depth, which is determined by the initial beam energy ($E_{beam}$).
This physical property enables highly localized dose delivery, allowing the treatment to be tailored so that the Bragg peak aligns with the tumor site. The ionization and thus therapeutic damage is concentrated within the malignant tissue, while sparing surrounding healthy cells.

This method critically depends on the accurate determination of the relative stopping power (RSP) of the charged particles used for irradiation. The RSP can be inferred from measurements of the particles' scattering angles and their residual kinetic energies after traversing a phantom ~\cite{raparia1997algebraicreconstructiontechniqueart}
, which is a simplified representation of target tissues or organs intended for irradiation.
\par
Unlike X-rays, the energy loss of charged particles cannot be inferred directly from intensity changes, as their paths are significantly affected by scattering. This means that traditional CT~\cite{hounsfield1973computerized} images result in unwanted uncertainties of the RSP, limiting the opportunities of the hadron therapy~\cite{solie2020image} for hadron therapy dosage planning. Within the Bergen pCT collaboration, our aim is to develop a dedicated particle detector system, together with algorithms capable of reconstructing charged-particle trajectories in order to accurately estimate both their residual kinetic energies and scattering angles~\cite{solie2020monte, Aehle_2023,pct_CNN}. 

In proton computed tomography, the first step involves directing charged particles of relatively high, {\cal O}(100) MeV energy onto the object. While these energies are modest compared to those used in high-energy physics (GeV–TeV scales), they are sufficiently high that the majority of the energy deposition occurs not within the phantom but in an external detector assembly.

In this work, we present two deep learning based algorithms designed to enhance charged particle identification within the pCT detector. The first algorithm reconstructs particle trajectories, while the second predicts the residual kinetic energy. Trajectory reconstruction is formulated as a bipartite matching problem, where the goal is to connect consecutive particle hits based on spatial proximity.

\section{Related works}
In our previous work~\cite{prevPCT}, we demonstrated that compact machine learning models can be used to reconstruct particle trajectories. Although individual matching performance was promising, the range of tests was considerably narrower than in the present study, and both matching quality and overall track reconstruction accuracy were substantially lower. The Kalman filter~\cite{KalmanFilter}, which is widely employed in high-energy physics~\cite{8708910,Lantz_2020}, is not suitable for pCT applications because of its high computational cost. Other approaches~\cite{lstmtracking,ParticleReconstruction} employ neural network architectures such as LSTMs or graph neural networks; however, these methods are typically developed for substantially higher particle energies and often rely on magnetic fields to guide scattering, conditions that do not 
apply in our experimental setup.
An iterative score based algorithm was introduced by the pCT collaboration, which relied on no deep learning usage ~\cite{helgetrack}.
Later a reinforcement learning based method was presented to be able to reconstruct particle trajectories  ~\cite{TobiasArticle}.

\section{Setup}
\begin{figure}[h!]
\begin{center}
    \includegraphics[width=1.\textwidth]{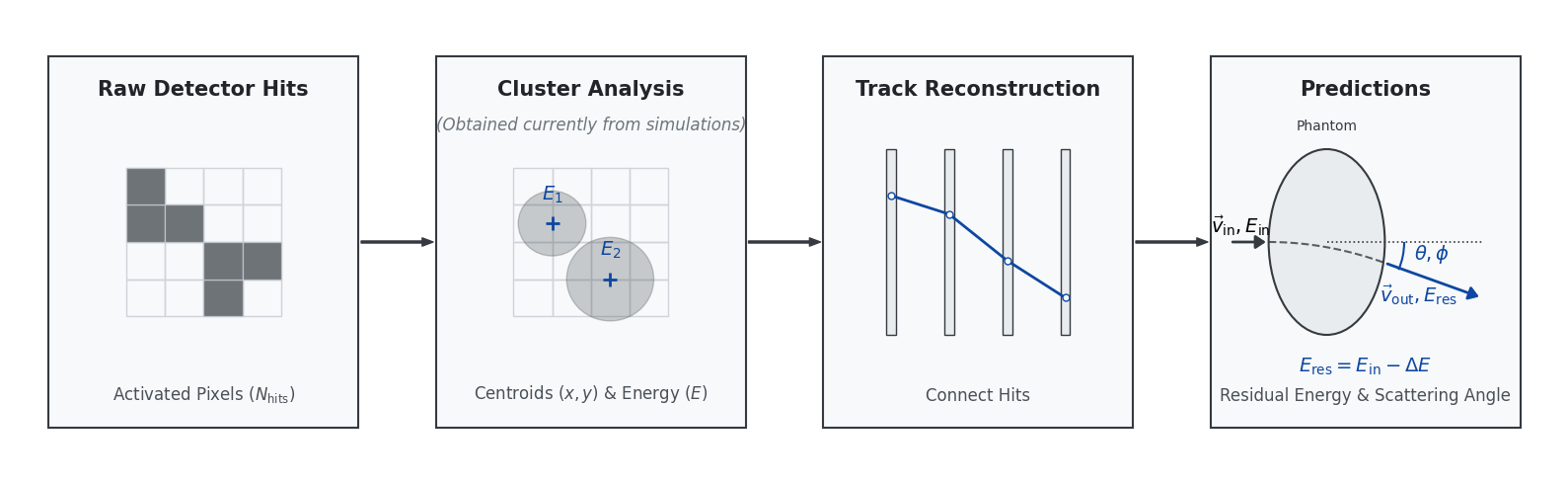}
    \caption{Step-by-step pipeline of detector data processing in the pCT pipeline. }
    \label{fig:track_pipeline}
\end{center}
\end{figure}

Our system follows a granular tracking-calorimeter design composed of two tracking layers behind the phantom, used to determine the scattering angles, followed by forty-one calorimetric layers, used to infer the incoming kinetic energy. 
Between the calorimetric layers, a 3.5 mm thick aluminum absorber~\cite{pctdetector} is placed to decelerate the particles, ensuring that they come to rest within the detector system. However, this material also introduces the possibility of inelastic scattering events. Each detector layer is constructed using ALICE ALPIDE sensors~\cite{AliceAlpide}, which provide binary information on whether a particle has traversed the pixel. From these sensor responses, we reconstruct the detector hits that are defined as the cluster of pixels~\cite{Pettersen2019,Alex_cluster}, the size
of which reflects the amount of energy deposited as the particle passes through. In Fig.~\ref{fig:track_pipeline} we present the pCT pipeline for detector analysis.

The dataset used in this study was generated with the OpenGATE~\cite{Gatetoolkit,g41,g42,g43,jan2004gate} (also called GATE) toolkit, a medical extension of Geant4~\cite{agostinelli2003geant4}, the widely used particle physics simulation framework. This simulation environment allows us to produce clean, fully controlled datasets that are well suited for supervised learning and systematic evaluation of our models. OpenGATE immediately produces the centroids and energies of the clusters. The pCT collaboration created the digital twin of the detector system, which we used for particle scattering simulations. The models were trained using water only phantoms with varying thicknesses. An experiment is defined as we beam in 10000 primaries into the  detector system and measure their respective detector information. Some of these particles stop in the phantom or suffer such scattering with such an angle that it prevents them entering the detector system. From the particles that enter the detector system we select a fixed amount (by default 100, except when we check the effect of this number), which reflects on how much particle can a detector layer register in one readout frame.  
The dataset was generated using water phantoms spanning 70–180 mm in 10 mm increments, with a fixed beam energy of 230 MeV, resulting in $\mathcal{O}(10^7)$ trajectories for each thickness.  
This has been split into 80\%-20\% to training and testing.
  For evaluation, the trained models were tested on both water phantoms and head phantoms in order to assess performance under different material compositions.

\section{Method}

If we denote the set of real tracks as $N_{real}= (n_1,n_2,..,n_k) $ and the set of reconstructed ones as $N_{rec} = (m_1,m_2,..,m_k)$, 
where $k$ represents the number of elements within a trajectory and $n_i,m_i$ represents the hits in detector layer $i$, the matching between the two can be defined as
\begin{equation}
\label{eq:match_acc}
    M(N_{real}= \{n_1,n_2,..,n_k\},N_{rec} = \{m_1,m_2,..,m_k\}) = \begin{cases}
    1 & \text{if } n_L=m_L \text{ for all } L = 1,...,k  \\
    0 & \text{otherwise},
    \end{cases}
\end{equation}
indicating that a trajectory is considered correctly reconstructed only if all corresponding elements are identical to their original counterpart. To ensure all tracks are of same length, we padded every track to the same size, but padding values are excluded from matching metric calculations. 

For image reconstruction we don't necessarily need the whole reconstructed track, but only certain key-points, which can be used to predict the residual kinetic energies and the scattering angles.
The latter can be determined from the hit positions in the first and second detector 
layers (the tracking layers, where no absorbent is added), whereas the kinetic energy is estimated from the energy depositions in the final detector layers. 
Incorporating these practical considerations, the matching criterion can be refined as

\begin{equation}
\label{eq:match_acc_mod}
M(N_{\text{real}}, N_{\text{rec}}) =
\begin{cases}
1 & \text{if } n_L = m_L \text{ for } L \in {1, 2, k-2, k-1, k} \\
0 & \text{otherwise},
\end{cases}
\end{equation}

meaning that a reconstructed trajectory is accepted as correct if it matches the real one at the first two and the last three spatial positions.

The motivation for using small deep learning algorithms comes from the necessity that the reconstruction of the trajectory of particles along with the generation of images should be done within a reasonable time  for medical use\cite{alme2020high}. Deep learning methods showed great performance in various areas compared to traditional algorithms \cite{lecun2015deep,krizhevsky2012imagenet}.

To address the limitations of traditional distance-based matching algorithms~\cite{hungarian_alg,Sinkhorn,Sinkhorn2,Ford_Fulkerson},which often fail under conditions of multiple Coulomb scattering and particle track crossings, as shown in our previous work\cite{prevPCT}, we employ a neural network that predicts the expected hit position (which is the center of the pixel cluster in the real detector system) in the adjacent detector layer based on the current location of the particle. This learned prediction serves as a physically informed prior, substantially reducing matching ambiguities caused by scattering-induced deviations and improving the robustness of the overall track reconstruction process.
The distance matrix in our method is calculated by equation~\eqref{eq:distmx},
\begin{equation}
\label{eq:distmx}
    S_{X_L,X_{L-1},X_{L-2}}= ||X_{L}-h_\theta(X_{L-1},X_{L-2})  ||^2,
\end{equation}
where $X_L$ represents the position vector in the $L$-th layer and $h_\theta$ represents the deep neural network that projects the position vector into the next layer. We have found no improvement extending input with further layer information.

The position prediction pipeline is based on a multilayer perceptron (MLP). We trained three separate neural networks with this identical MLP architecture (linear layers with 8700 parameters from 3 hidden layers) allowing for different weights to account for variations in the geometry of the first two detector layers: tracking-tracking-calorimetric, tracking-calorimetric-calorimetric and calorimetric-calorimetric-calorimetric.
We applied the same neural network for all the calorimetric layers, as their physical configurations are uniform. 

\par
For the residual kinetic energy estimation we built another MLP with 2 hidden layers and LeakyReLU activation. The last position of each particle, the last detector layer they hit, is already a good first order estimate of what their energy was before entering the detector system. Some particles leave the detector system before they stop; therefore,  we do not want to use them for energy estimation. We want to focus on particles that stop in the detector system, so we had to apply a filter for every track. First, we define an expected Bragg-peak layer, or target layer, by equation~\eqref{eq:targetlayer},  
\begin{equation}
\label{eq:targetlayer}
    L_t = argmax(P(x_L)),
\end{equation}
where $x_L$ denotes the number of particles stopping at layer $L$  and $P(x_L)$ denotes the (empirical) probability distribution of $x_L$. 

Equation~\eqref{eq:targetlayer} identifies the layer $L_t$ where the majority of particles come to rest, thereby approximating the location of the Bragg peak. Given that the position of the Bragg-peak shifts by one calorimetric layer for each increase of 8~MeV in the residual kinetic energy for our setup, $L_t$ alone allows for the determination of the residual kinetic energy with a precision of $\pm 4$~MeV.  
Then to further improve this prediction, we use the approximated Bragg peak  to normalize energy depositions:

\begin{equation}
    \hat{E}_{L_t \pm1} = E_{L_t \pm 1}/E_{L_t},
\end{equation}
where $E_{L_t}$ is the energy deposited at the detector layer $L_t$  and $E_{L_t \pm 1}$ is the energy deposited at the neighboring layers. Finally, we feed these values into our compact neural network:
\begin{equation}
    \hat{E} = g_{\theta}(L_t,\hat{E}_{L_t-1},\hat{E}_{L_t+1})
\end{equation}
where $\hat{E}$ is the predicted residual kinetic energy and $g_{\theta}$ is our neural network, parameterized with $\theta$.

\section{Results}
\subsection{Matching}
\begin{figure}[h!]
\begin{center}
    \includegraphics[width=0.65\textwidth]{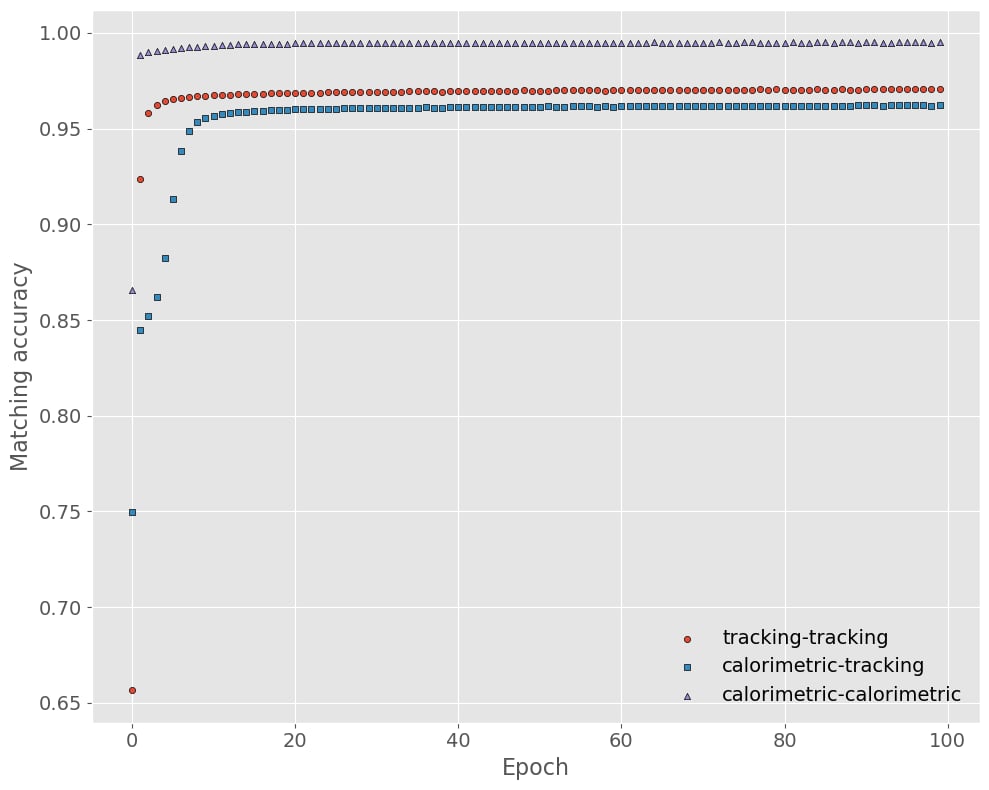}
    \caption{Matching accuracy between subsequent trajectory points in the validation dataset of the pCT Detector system.}
    %\PgComment{a tengelyeken a számokat még egy kicsit kéne növelni, és a legend betűméretét is növeld nyugodtan, van hely.}
    \label{fig:val_acc}
\end{center}
\end{figure}

First, we evaluated the layer-layer matching capabilities of our model, which is just evaluated between two adjacent layer and not for the whole track.
  It is shown in Figure ~\ref{fig:val_acc}  that an accuracy ( defined as $\frac{N_{correct}}{N_{all}}$, where $N_{correct}$ is the correctly matched hits and $N_{all}$ is all the hits) of at least 95\% is achievable, while over 99.5\% accuracy is obtained within the calorimetric layers. We evaluated our methods for 100 trajectories per frame, i.e. considering 100 trajectories at the same time.
We additionally compared our approach against several standard matching algorithms commonly employed in particle tracking. The baselines included greedy (nearest neighbor) matching, the Hungarian algorithm~\cite{hungarian_alg}, Sinkhorn-based optimal transport~\cite{Sinkhorn,Sinkhorn2}, and the Ford–Fulkerson method~\cite{Ford_Fulkerson}. As summarized in Table~\ref{table:matching}, in the tracking–tracking configuration (Layer 0–1), both our method and Ford–Fulkerson exceed 91\% accuracy however, our model consistently provides higher accuracy with substantially greater confidence and orders of magnitude less time. A similar trend is observed in the tracking–calorimetric transition (Layer 1–2), where our method again outperforms all baselines. In contrast, the differences between all methods become less significant in the calorimetric–calorimetric region (Layer 2–3). This is expected, since the distance between detector layers is much smaller and the number of particles (primaries) decreases as protons either stop or leave the detector-system, making it easier to distinguish between data points.

\begin{table}[htbp]
\caption{Performance of different matching algorithms between subsequent trajectory points. Best results are in bold.}
\label{table:matching}
\centering
\begin{tabular}{l c c c}
\hline
Matching method & Layer 0-1 matching & Layer 1-2 matching & Layer 2-3 matching \\
\hline
Our method
& $\mathbf{97.20 \pm 0.29}$
& $\mathbf{95.84 \pm 0.36}$
& $\mathbf{98.42 \pm 0.26}$ \\

Greedy matching
& $52.30 \pm 2.08$
& $71.88 \pm 1.42$
& $97.95 \pm 0.30$ \\

Hungarian algorithm
& $74.59 \pm 1.28$
& $86.50 \pm 1.33$
& $96.82 \pm 0.67$ \\

Sinkhorn matching
& $70.96 \pm 1.86$
& $87.69 \pm 0.94$
& $98.38 \pm 0.29$ \\

Ford-Fulkerson algorithm
& $91.59 \pm 3.55$
& $93.21 \pm 3.05$
& $\mathbf{98.42 \pm 1.69}$ \\
\hline
\end{tabular}
\end{table}

\par
\begin{figure}[h!]
    \centering
    \includegraphics[width=.32\textwidth]{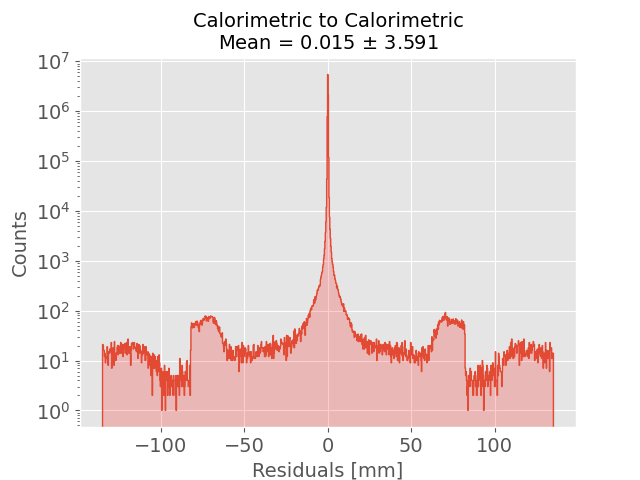}
    \includegraphics[width=.32\textwidth]{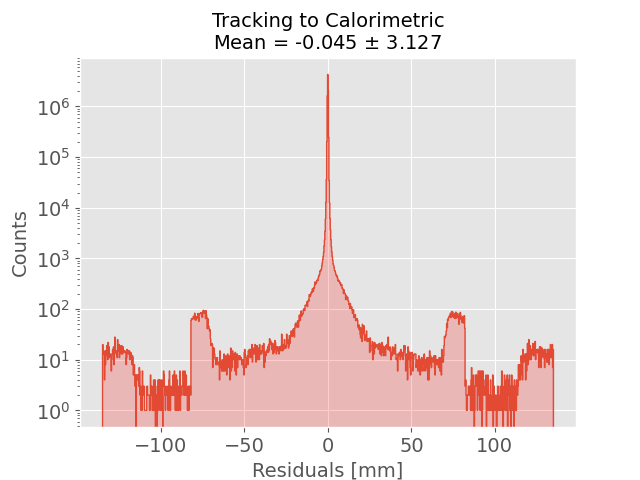}
    \includegraphics[width=.32\textwidth]{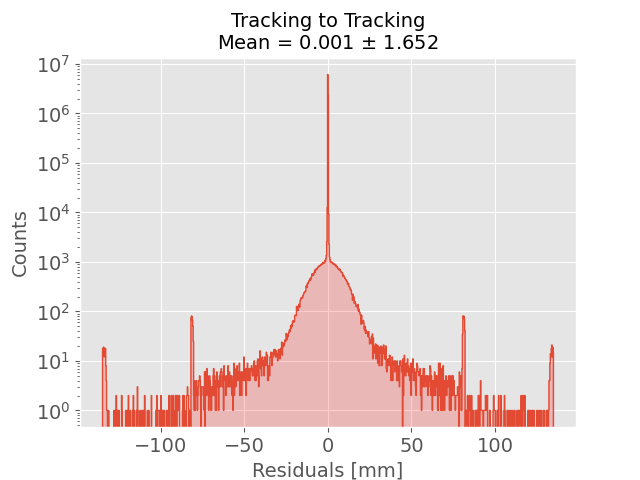}

    \caption{Difference between the true positions and the predicted positions (mm) from our model in logarithmic scale. Results on predictions from tracking layer to tracking layer  (left), predictions from calorimetric layer to tracking layer (middle), predictions from calorimetric to calorimetric layer (right).
    }
    \label{fig:residuals}
\end{figure}

We can see in Figure ~\ref{fig:residuals}, that the residual plots are sharply skewed around 0, meaning that the predictions are approximated very closely. However, there are cases when the  predictions deviates strongly, and we will address them later in the pipeline.

\subsection{Tracking}
Moving beyond layer-to-layer hit matching, we evaluated the performance of the track reconstruction pipeline using the criterion defined in Eq.~(\ref{eq:match_acc_mod}). Under this metric, the proposed method achieved an accuracy of 75.09\%. 
To further improve this performance, we excluded trajectories that exhibit inconsistent matching behavior. Specifically, a match between detector layers was accepted only if the optimal assignment was mutually consistent in both directions of the distance matrix. This criterion can be formulated as

\begin{equation}
\label{eq:maskuncertain}
j = \operatorname*{argmax}_{k} S_{ik}
\quad \text{and} \quad
i = \operatorname*{argmax}_{l} S_{lj},
\end{equation}

where $S_{ik}$ denotes the matching probability between hit $i$ in layer $X_{L-1}$ and hit $k$ in layer $X_{L}$. Trajectories violating bidirectional consistency condition, $j$ and $i$ does not correspond for the same hit, are classified as uncertain and excluded from further reconstruction. In practice, this criterion removes the hits associated with uncertain trajectory reconstructions, thus improving the overall consistency of the track reconstruction. 
 
\begin{figure}[h!]
\begin{center}
\includegraphics[width=0.99\textwidth,height=80mm]{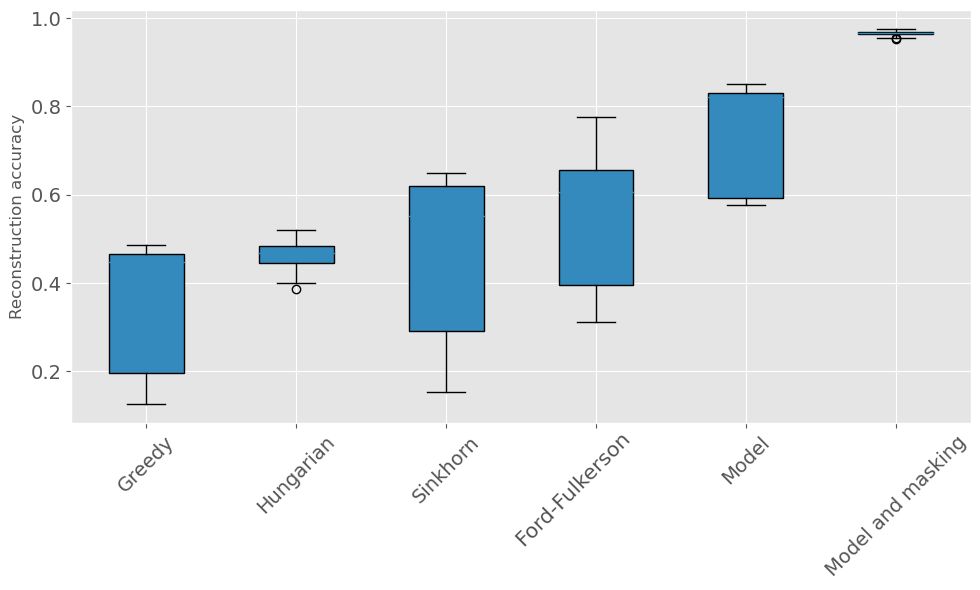}
    \caption{Reconstruction accuracy of full trajectories without \eqref{eq:maskuncertain} masking (except the second column), using different matching algorithms.} 
    \label{fig:masked_acc}
\end{center}
\end{figure}

\begin{figure}[h!]
    \centering
    \includegraphics[width=.36\textwidth]{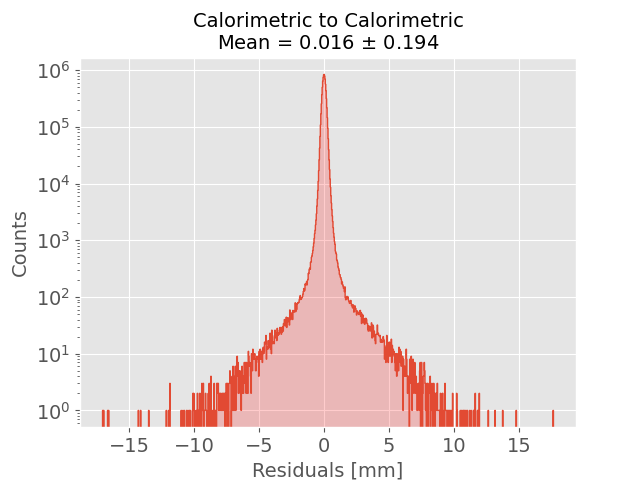}
    \hspace*{-10mm}
    \includegraphics[width=.36\textwidth]{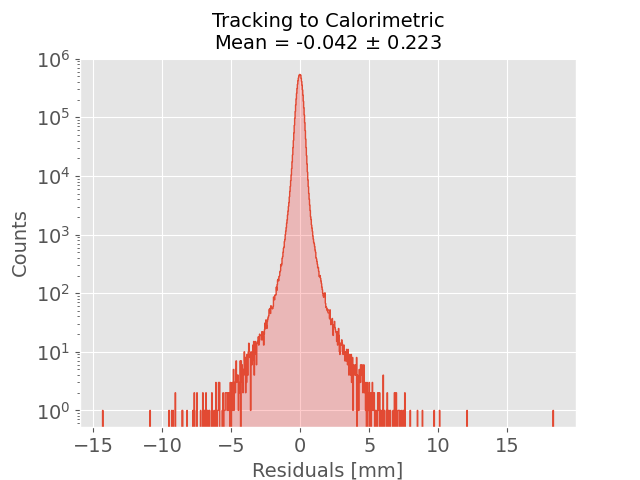}
    \hspace*{-10mm}
    \includegraphics[width=.36\textwidth]{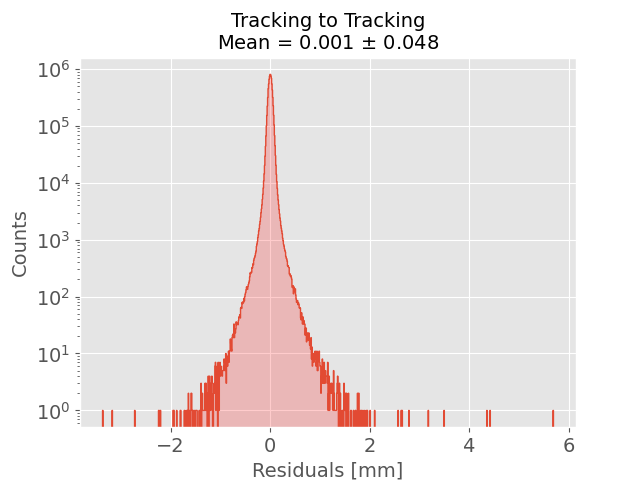}
    \caption{Difference between the true positions and the predicted positions(mm) with the the consistency filtered method on logarithmic scale. Results on predictions from tracking layer to tracking layer  (left), predictions from calorimetric layer to tracking layer (middle), predictions from calorimetric to calorimetric layer (right).
    }
    \label{fig:residuals_filtered}
\end{figure}

 Applying this consistency criterion for 100 protons in the readout frame (p/F), we have filtered out almost 50\% of the trajectories however, the accuracy has increased to $\approx 92.33\%$.
 As illustrated in Fig.~\ref{fig:masked_acc}, although the algorithm without the consistency condition achieves a reasonably high average performance, the quality of the final results exhibits considerable variability. After applying the consistency filter Eq. ~\eqref{eq:maskuncertain}, not only does the mean performance increase, but the overall consistency of the predictions is also enhanced significantly with highly reduced spread (see Fig.~\ref{fig:residuals_filtered}).

Fig.~\ref{fig:masked_acc} demonstrates that our method, combined with uncertainty-based masking, consistently outperforms all other matching strategies in the full track reconstruction pipeline. Applying the same masking strategy to the classical algorithms does increase their accuracies too, however, this improvement comes at the cost of discarding a large fraction of particle trajectories. Such an aggressive dropout is inappropriate for our image reconstruction algorithms.
In contrast, our model achieves a favorable balance between accuracy gains and data retention, making it substantially more suitable for practical reconstruction workflows.

Table~\ref{tab:particle_performance} demonstrates that the proposed reconstruction pipeline maintains high accuracy even as the number of trajectories per frame increases. Although higher particle densities introduce additional track ambiguities, the degradation in reconstruction accuracy remains moderate.
\begin{table}[htbp]
    \centering
    \caption{Matching accuracy and fraction of dropped tracks with different number of protons in the frame.}
    \label{tab:particle_performance}
    \begin{tabular}{@{} c  c c @{}}
    \hline
        Initial trajectories & Accuracy (\%) & Drop Fraction (\%) \\
        \hline
        200 & 93.2 $\pm$ 0.5 & 58.1 $\pm$ 2.5 \\
        150 & 94.9 $\pm$ 0.5 & 54.1 $\pm$ 2.8 \\
        100 & 96.5 $\pm$ 0.5 & 47.5 $\pm$ 3.1 \\
        50 & 98.3 $\pm$ 0.4 & 34.7 $\pm$ 3.2 \\
    \hline
    \end{tabular}
\end{table}

The models were trained and validated primarily on water phantoms to systematically characterize the impact of material thickness on reconstruction performance. 
Evaluating the performance of our model by applying the strict matching condition - ratio of tracks where every hit was matched correctly to all tracks - from Ref.~\cite{TobiasArticle} to all vertices, the accuracy decreases slightly to 92.1\% ($p/F=50$), 90.8\% ($p/F=100$), 89.9\% ($p/F=150$), and 89\% ($p/F=200$). For comparison, we cite the average accuracies reported in the previously mentioned article with reinforced learning for the water phantom: 93.5\% (50), 88\% (100), 83.2\% (150), and 79\% (200). 

Applying the model (trained on water phantom) to the head phantom, we observed a degradation in accuracy of approximately 5\%, whereas Ref.~\cite{TobiasArticle} reported a degradation of only about 1\%.
\subsection{Energy prediction}

The model used for predicting kinetic energies after phantom traversal (with fixed 230 MeV initial Kinetic energy) was evaluated with different water phantom thicknesses from range of 110 mm to 200 mm with a 10 mm step size. The residual energy (energy of the particles after they exit the phantom) is logged by the OpenGate simulation and then approximated by our neural network.

\begin{figure}[h!]
    \centering
    \includegraphics[width=.40\textwidth]{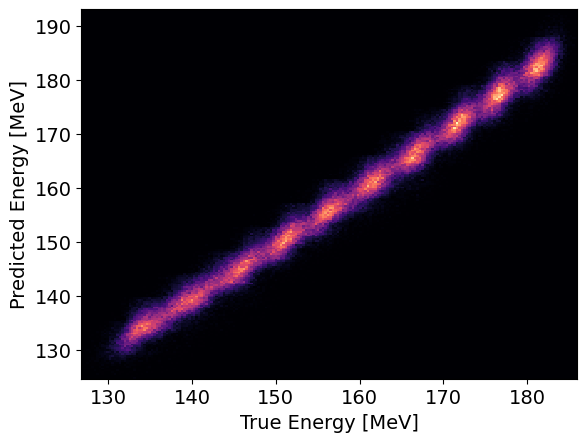}
    \includegraphics[width=.49\textwidth]{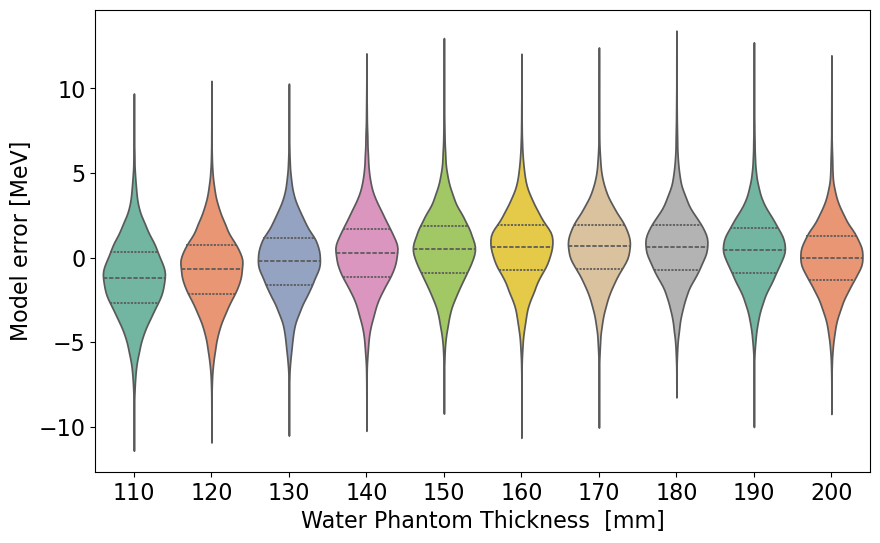}
    \caption{The 2 dimensional histogram of the true values and the predicted values(left panel) and the error of the prediction in different initial kinetic energies (right panel). Statistics are calculated over 1000 runs, where every run contained 200 trajectories.
    }
    \label{fig:ener_preds}
\end{figure}

As shown in the right panel of Fig.~\ref{fig:ener_preds}, the mean error is centered around zero. The left panel of Fig.~\ref{fig:ener_preds} further illustrates that predictions follows the same pattern as the ground truth energies with some prediction noise. 

The width of the violins on the right panel of \ref{fig:ener_preds} corresponds to the relative frequency of a given error value, showing that most prediction errors are concentrated around 0 MeV. Colors are included only for visual distinction between the individual distributions. The overall energy approximation accuracy of our model was 1.2 MeV.

\subsection{Performance}
A critical aspect of proton computed tomography is the computational speed of the reconstruction process. For clinical applications, both the track reconstruction algorithm and subsequent image reconstruction must operate within a short time frame to enable timely treatment initiation. Since our methods are based on deep learning, they benefit from significant acceleration through GPU utilization. Furthermore, during track reconstruction, it is unnecessary to iterate through the entire detector system, as protons typically traverse only a limited number of layers.

\begin{figure}[h!]
\begin{center}
    \includegraphics[width=0.5\textwidth]{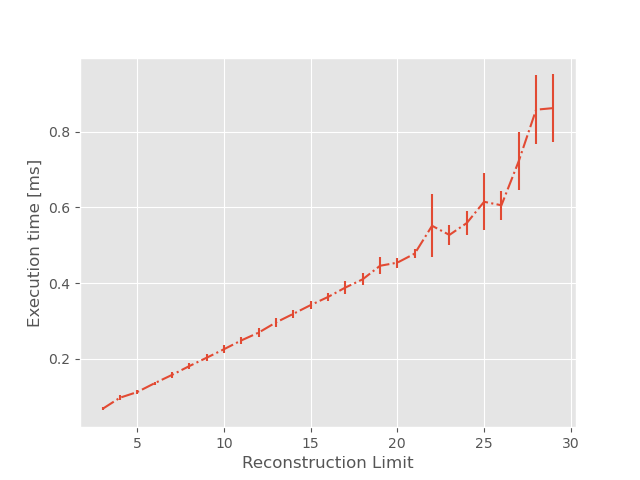}
    \caption{Trajectory reconstruction time of a single particle in the function of the reconstruction limit. Averaged out over the whole test dataset.
    }
    \label{fig:rectime}
\end{center}
\end{figure}

To analyze the relationship between reconstruction depth and processing time, we measured the algorithm’s execution time for varying reconstruction limits defined as the final detector layer included in the reconstruction over 1000 runs. While the actual reconstruction limit in real life scenario depends on the maximum penetration depth of individual particles, this parameterization allows consistent and controlled benchmarking. As shown in Fig.~\ref{fig:rectime}, even for longer tracks extending to the 30th detector layer, the reconstruction time per  processed trajectory remains below 1 ms on NVIDIA TESLA T4 GPUs.

% TIME

\section{Discussion}

In this study, we proposed a fast and accurate data processing pipeline for proton computed tomography, focusing on particle trajectory reconstruction and kinetic energy estimation using compact deep learning models. The results demonstrate that small neural networks, when integrated with physically informed features, can overcome the limitations of traditional distance-based matching algorithms in scenarios with substantial multiple Coulomb scattering and track crossings. By predicting expected hit positions instead of relying solely on geometric proximity, our method significantly increases matching stability with high layer to layer accuracies and global track reconstruction performance.

The energy estimation module demonstrated a mean error of approximately 1.2 MeV. The use of physically motivated features, such as the expected Bragg-peak layer and normalized energy depositions from neighboring layers, proved particularly effective for achieving this level of precision. These findings suggest that relatively simple neural network architectures, when combined with domain-specific priors, are capable of delivering performance comparable to more complex models while offering considerable advantages in interpretability and computational speed.

From a computational standpoint, the proposed pipeline shows promising potential for integration into clinical workflows. Deep learning inference enables rapid processing, with track reconstruction times remaining below 1 ms even for extended trajectories.  This efficiency aligns with the time requirements of image generation in proton therapy, where delays must be minimized to ensure accurate dose delivery. Furthermore, the pipeline does not rely on a fixed proton/frame value, it is dynamically adjusted.

Despite these strengths, several limitations warrant discussion. First, the models were trained and evaluated on data generated with the OpenGate/Geant4 simulation framework, which provides clean, well-characterized ground truth. 
Although this enables controlled experimentation, real detector data introduce additional complexities, including noise, sensor inefficiencies, and potential geometric misalignment. In our project we used simulation data, but in real life applications the detector hits are constructed as the clusters of active pixels, which can introduce some more inaccuracies in the pipeline. Future work will therefore focus on applying and validating the pipeline on experimental pCT acquisitions from the Bergen collaboration. Second, while the use of separate neural networks for distinct detector configurations proved effective, a unified conditioned model could potentially improve scalability and simplify deployment, exploring such architectures remains an avenue for future investigation.
Overall, this work demonstrates that compact neural architectures, when combined with physically informed features and uncertainty handling, provide an effective and computationally efficient solution for particle tracking and energy estimation in proton computed tomography. These findings represent a meaningful step toward real-time, clinically deployable pCT reconstruction pipelines, and establish a foundation for future refinement using experimental datasets and more advanced model architectures.

\section{Conclusion}
We developed a comprehensive data processing pipeline for proton computed tomography (pCT), consisting of two major components. The first is a tracking module, comprising an ensemble of compact neural networks designed to predict adjacent detector hit positions across the various configurations of the pCT detector system. This approach substantially improves upon conventional distance-based matching methods, achieving exceptionally high matching accuracies exceeding 95\% across all detector layer configurations.

To further enhance robustness, we introduced a consistency filtering procedure that removes mismatched hits, thereby improving the overall tracking precision. Additionally, by employing a physically interpretable evaluation metric, we achieved accuracies above 96.5\% (100 p/F) on the whole particle trajectory reconstruction. The parallelism of deep learning inference enabled the evaluation time of the tracking module to remain suitable for medical application.

\section*{Acknowledgements}
This research is supported by the NKFIH DKOP-23 Doctoral Excellence Program of the Ministry for Culture and Innovation,  the NKKP ADVANCED\_25-153456, 2021-4.1.2-NEMZ\_KI-2024-00058,2025-1.1.5-NEMZ\_KI-2025-00005 and 2025-1.1.5-NEMZ\_KI-2025-00013.
The authors acknowledge the research infrastructure provided by the Hungarian Research Network (HUN-REN) and the Wigner Scientific Computing Laboratory. Authors G. B., G. P. and B. D. were supported by the European Union project RRF-2.3.1-21-2022-00004 within the framework of the Artificial Intelligence National Laboratory.

\bibliographystyle{vancouver} 
\bibliography{pCT_tracking_BP}

\end{document}